\documentclass[%
preprint,
groupedaddress,
 amsmath,amssymb,
 aps,
showpacs
]{revtex4-1}

\usepackage{graphicx}
\usepackage{dcolumn}
\usepackage{bm}
\usepackage[bookmarkstype=toc,colorlinks=true,urlcolor=blue,linkcolor=blue,citecolor=blue,linktocpage=true,bookmarks=true,setpagesize=false]{hyperref}

\usepackage{braket}

\def\rmi{{\rm i}}
\allowdisplaybreaks[4]
\def\e{\epsilon}

\begin{document}


\title{Analysis of the Pancharatnam--Berry phase of vector vortex states \\using the Hamiltonian based on the Maxwell--Schr\"odinger equation}

\author{Masato Suzuki}
\affiliation{Department of Applied Physics, Hokkaido University, Kita-13, Nishi-8, Kita-ku, Sapporo 060-8628, Japan}
\author{Keisaku Yamane}
\affiliation{Department of Applied Physics, Hokkaido University, Kita-13, Nishi-8, Kita-ku, Sapporo 060-8628, Japan}
\author{Kazuhiko Oka}
\affiliation{Department of Applied Physics, Hokkaido University, Kita-13, Nishi-8, Kita-ku, Sapporo 060-8628, Japan}
\author{Yasunori Toda}
\affiliation{Department of Applied Physics, Hokkaido University, Kita-13, Nishi-8, Kita-ku, Sapporo 060-8628, Japan}
\author{Ryuji Morita}
\email{morita@eng.hokudai.ac.jp}
\affiliation{Department of Applied Physics, Hokkaido University, Kita-13, Nishi-8, Kita-ku, Sapporo 060-8628, Japan}

\date{\today}

\begin{abstract}
We derived the Berry connection of vector vortex states (VVSs) from the ``true'' Hamiltonian obtained through the Maxwell--Schr\"odinger equation for an inhomogeneous anisotropic (IA) medium, and we experimentally demonstrated measurement of the corresponding Pancharatnam--Berry (PB) geometrical phase of VVSs. The PB phase (PBP) of VVSs can be divided into two phases: homogeneous and inhomogeneous PBPs. Homogeneous and inhomogeneous PBPs are related to the conventional PBP and the spatially-dependent geometric phase given by an IA medium such as a polarization converter, respectively. 
We theoretically detected that inhomogeneous PBP accumulation originates from the gauge dependence of the index of the hybrid-order Poincar\'e sphere, which provides an alternate method for understanding optical spin--orbital angular momentum conversion. The homogeneous PBP, which is explicitly observed for the first time, has implications for quantum state manipulation and information processing.
\end{abstract}

\pacs{03.65.Vf, 42.25.Ja, 42.50.Tx}
\maketitle


\textit{Introduction.}---The Pancharatnam--Berry phase (PBP) is a geometrical phase \cite{anandan1992geometric} associated with polarization of light 
 \cite{P1st}. The PBP has been experimentally observed using a homogeneously distributed polarization light state \cite{PhysRevLett.60.1211,*Chyba:88,*Visser:10}. Recently, using \textit{spin--orbit} converters \cite{PhysRevE.60.7497}, Milione \textit{et al.} \cite{PhysRevLett.108.190401} conducted a pioneering exploration of PBPs for vector vortex states (VVSs). Here, the VVSs, having received  attracting attention for many applications \cite{PhysRevLett.110.143603,*parigi2015storage}, are light states having both inhomogeneous phase and polarization distributions \cite{1367-2630-9-3-078}. 
On a higher-order Poincar\'e sphere (HiOPS), which is the space of higher-order Stokes parameters (HiOSPs) \cite{PhysRevLett.107.053601}, Milione \textit{et al.} show pathways made by the \textit{spin--orbit} converters. 

A polarization converter to generate a vector vortex beam can make a path from a state on one (HiOPS) to a state on another HiOPS \cite{Holleczek:11}, whose process can be described on a hybrid-order Poincar\'e sphere (HyOPS) \cite{PhysRevA.91.023801}. Yi \textit{et al.} proposed not only the HyOPS but also the PBP for the HyOPS, which are acquired similarly to the method of Milione \textit{et al.}  \cite{PhysRevLett.108.190401}.

Milione \textit{et al.} and Yi \textit{et al.} obtained the PBP from the Berry connection \cite{Berry45}. However, neither of these studies explicitly showed the Hamiltonian, and the wavevectors used to express states on a HiOPS or a HyOPS are seemingly invalid. These invalid wavevectors lead to the inaccurate conclusion that the PBP did not correspond to the adiabatic polarization state change on a HiOPS nor a HyOPS. Being not able to be expressed by these invalid wavevectors, the intermediate states in the \textit{spin--orbit} converters are, hence, not described on a HiOPS or a HyOPS but on the spheres for a \textit{spin--orbit} converter \cite{Suzuk_prlcomment}. Moreover, it is difficult to justify the Hamiltonian for an optical system \cite{PhysRevLett.57.933}. Although Berry has shown the procedure to acquire the PBP for a uniformly polarized light state \cite{doi:10.1080/09500348714551321}, the Hamiltonian is just the density matrix of the circularly polarized states, which does not provide an appropriate equation of motion
for wave plates. Therefore, it is essential to acquire the true PBP from the true Hamiltonian describing an inhomogeneous anisotropic (IA) medium such as $q$-plates \cite{PhysRevLett.96.163905}, which is one of typical \textit{spin--orbit} converters.

In the present letter, in order to accurately discuss the PBP for VVSs in the right way, we will reestablish the Berry connections of VVSs on the HiOPS and HyOPS. We first acquire the ``true'' Hamiltonian of a $q$-plate as an extension of Refs.~\cite{PhysRevLett.80.1888,*PhysRevE.81.036602}, which acquire the Hamiltonian of  homogeneous birefringent media from the Maxwell--Schr\"{o}dinger equation. 
Furthermore, we will experimentally measure the PBP for VVSs and demonstrate it to be a ``true'' PBP given by IA media.

\textit{Hybrid-order Stokes parameters.}---We use bra-ket notation to describe a VVS:
\begin{eqnarray}
	\ket{\psi} &=& \frac{\psi_{+,l}e^{-\rmi l_+\phi}}{\sqrt{2}}\begin{pmatrix}1\\\rmi\end{pmatrix} + \frac{\psi_{-,l}e^{\rmi l_-\phi}}{\sqrt{2}}\begin{pmatrix}1\\-\rmi\end{pmatrix}\nonumber\\
	&=& \frac{e^{\rmi l'\phi}}{\sqrt{2}}\left [\psi_{+,l}e^{-\rmi l\phi}\begin{pmatrix}1\\\rmi\end{pmatrix} + \psi_{-,l}e^{\rmi l\phi}\begin{pmatrix}1\\-\rmi\end{pmatrix} \right ],\label{phidef000}
\end{eqnarray}
where $\phi$ is the azimuthal angle, $\psi_{\pm,l_\pm}$ are amplitude functions, $l_+$ and $l_-$ indicate topological charges of the left- and right-circularly polarized states, respectively, and $l'(=(-l_++l_-)/2)$ and $l(=(l_++l_-)/2)$ are azimuthal indices.
We require the inner product of the vector to be unity ($|\psi_{+,l}|^2+|\psi_{-,l}|^2=1$). By definition, the hybrid-order Stokes parameters (HyOSPs) are described as 
\begin{equation}
	 \bm{S}^{-l_+,l_-} = \left [1,S_1^{-l_+,l_-},S_2^{-l_+,l_-},S_3^{-l_+,l_-} \right ]^\mathrm{T}
	=\braket{\phi|\bm{\sigma}|\phi},
\end{equation}
where $\bm{\sigma}=[\sigma_0(\equiv\hat{1}),\sigma_1,\sigma_2,\sigma_3]^\mathrm{T}$ are the Pauli spin matrices \cite{Berry45}, 
and $\ket{\phi}$ represents $(\psi_{+,l},\psi_{-,l})^\mathrm{T}$. Since $S_i^{-l_+,l_-} \!=\! S^l_i\,(i\!=\!1,2)$ and $S_3^{-l_+,l_-} \!\equiv\! S_3$, we hereafter express $\bm{S}_{-l_+,l_-}$ as $\bm{S}_l$, and $\tilde{\bm{S}}_l$ as $\left [S^l_1,S^l_2,S_3 \right ]^\mathrm{T}$. If $l$ is an integer, $\bm{S}_l$ represents the HiOSPs.

\textit{Hamiltonian for $q$-plates.}---Following Refs.~\cite{PhysRevLett.80.1888,PhysRevE.81.036602}, the Maxwell--Schr\"odinger equation for an IA medium, whose transverse dielectric tensor is described by $\hat{\epsilon}_\bot = (\epsilon_{ij})\,(i,j=x,y)$, is
\begin{equation}
	2\rmi\sqrt{\epsilon}k^{-1}\partial_z \ket{\phi} = (\epsilon\hat{1}-T_l\hat{\epsilon}_\bot T_l^\dag)\ket{\phi} \label{eq:peqH00},
\end{equation}
where $\epsilon$ is a dielectric value, $k$ is a wave number, and 
\begin{equation}
T_l=\begin{pmatrix}e^{\rmi l \phi}&-\rmi e^{\rmi l\phi}\\ e^{-\rmi l\phi}&\rmi e^{-\rmi l\phi}\end{pmatrix}\nonumber
\end{equation}
is a transform matrix from the $x,y$-basis representation $\ket{\psi}$ to the circularly polarized optical vortex basis representation $\ket{\phi}$ \cite{supp01}. 
Here, the Hamiltonian $\mathcal{H}_l$ is given by $ \epsilon\hat{1}-T_l\hat{\epsilon}_\bot T_l^\dag$. 

The transverse relative permittivity tensor of a $q=l, \alpha_0=l\bar{\alpha}$ wave plate is described using a rotational matrix $R_\theta$:
\begin{equation}
	\hat{\epsilon}_\bot = R_{l(\phi+\bar{\alpha})}\begin{pmatrix}\epsilon_\mathrm{o}&0\\0&\epsilon_\mathrm{e}\end{pmatrix}R_{-l(\phi+\bar{\alpha})}.
\end{equation}
Note that the definitions of $q$ and $\alpha_0$ are given in Ref. \cite{PhysRevLett.96.163905}, and when $l\!=\!0$, $l\bar{\alpha}$ is replaced with $\bar{\alpha}$. 
Hence, the Hamiltonian is calculated to be
$\mathcal{H}_l = \epsilon_-\left (\cos(l\bar{\alpha})\sigma_1 + \sin(l\bar{\alpha})\sigma_2 \right ),$
where $\epsilon_\pm = \pm(\epsilon_\mathrm{o}\pm\epsilon_\mathrm{e})/2$ and  $\epsilon=\epsilon_+$. The evolution equation for $\tilde{\bm{S}}_l$ is given \cite{PhysRevE.81.036602} by 
\begin{equation}
\frac{\mathrm{d}\tilde{\bm{S}}_l}{\mathrm{d}\delta} = \tilde{\bm{S}}_l\times \left [-\cos(2l\bar{\alpha}),-\sin(2l\bar{\alpha}),0\right ]^\mathrm{T}, \label{eq:evo_eq00}
\end{equation}
where $\delta=\delta(z) = k\epsilon_-z/\sqrt{\epsilon_+}$ is the retardance phase. This equation supplies the true path made by a $q$-plate because Eq.~\eqref{eq:evo_eq00} represents precession on the HyOPS. 

\textit{Berry connection of vector vortex states.}---We now obtain the PBP of VVSs through the Berry connection. Since $\mathcal{H}_l$ is equivalent to a spin-1/2 system Hamiltonian, the Berry connection is expressed by
\begin{equation}
\bm{A}(\tilde{\bm{S}}_l) = -\bm{e}_{\zeta_l} (2R_l)^{-1}\tan\frac{\xi_l}{2} + \nabla_{\tilde{\bm{S}}_l} \Phi(\tilde{\bm{S}}_l), \label{eq:vecpot00}
\end{equation}
where $\Phi(\tilde{\bm{S}}_l)$ is a scalar potential \cite{Holstein89}. 
Here, we used the spherical coordinates $\tilde{\bm{S}}_l = R_l\bm{e}_{R_l} + \xi_l\bm{e}_{\xi_l} + \zeta_l\bm{e}_{\zeta_l}$, where $R_l$, $\xi_l$ and $\zeta_l$ are the radial distance, polar angle, and azimuthal angle in the $l$th HyOPS, and $\bm{e}_i\,(i=R_l,\xi_l,\zeta_l)$ represents the unit vector for the $i$ axis. 
We define the PBP of VVSs through the Berry connection as 
\begin{equation}
\gamma_\mathrm{Berry} = \sum_{i=0}^{i_\mathrm{max}} \int_{\tilde{\bm{S}}_{l_i}^\mathrm{S}}^{\tilde{\bm{S}}_{l_i}^\mathrm{F}} \bm{A}(\tilde{\bm{S}}_l)\cdot \mathrm{d}\tilde{\bm{S}}_{l_i},
\end{equation}
where we require a closed loop in the general meaning.  $\tilde{\bm{S}}_{l_0}^\mathrm{S}=\tilde{\bm{S}}_{l_\mathrm{max}}^\mathrm{F}$ and $\tilde{\bm{S}}_{l_i}^\mathrm{S}=\tilde{\bm{S}}_{l_{i-1}}^\mathrm{F}$, but $l_0\neq l_\mathrm{max}$ and $l_i\neq l_{i-1}$ are accepted if $\tilde{\bm{S}}_{l_i}^\mathrm{S}=[0,0,\pm 1]^\mathrm{T}$. 

\begin{figure}[tb]
\includegraphics[width=0.4\textwidth]{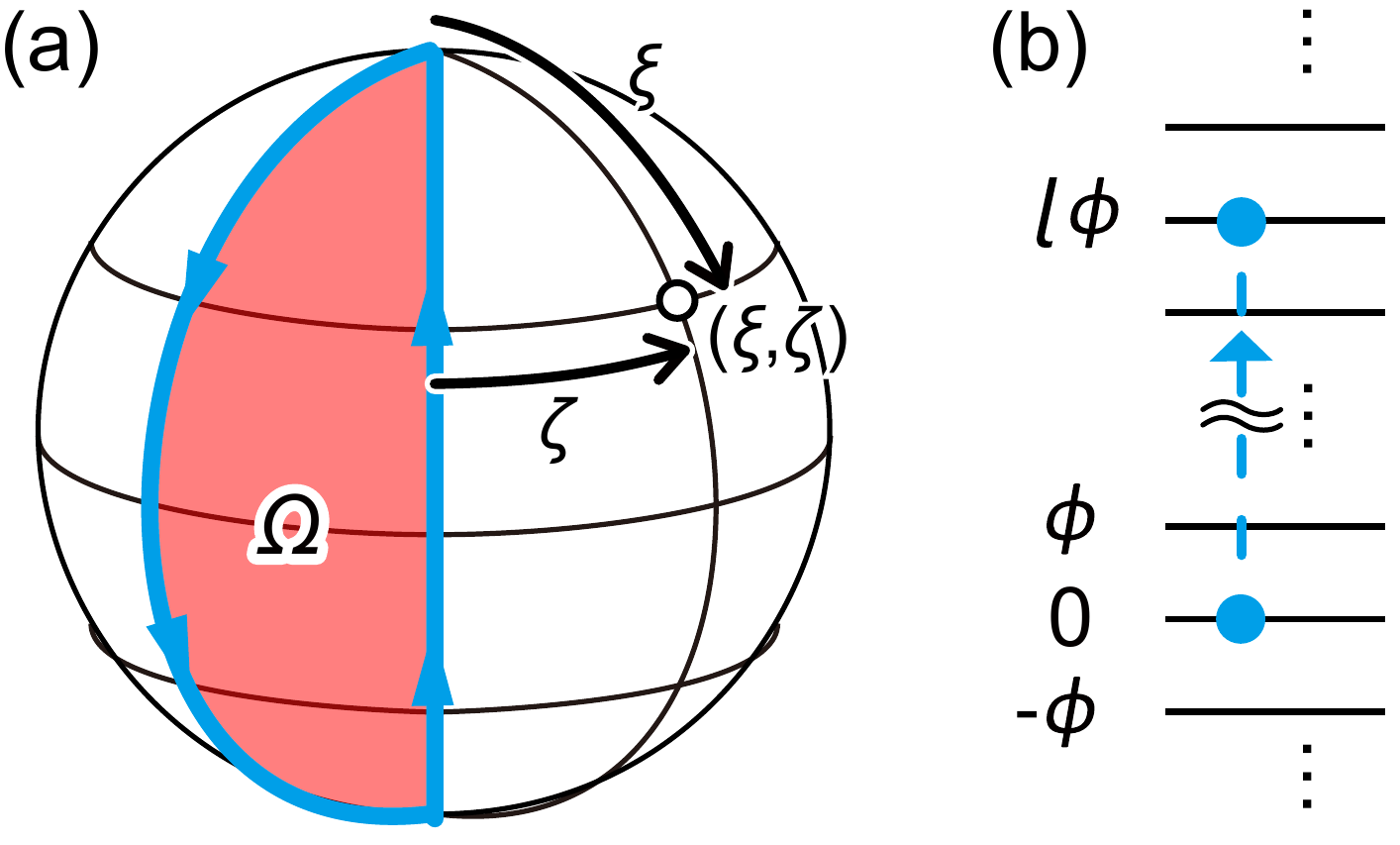}
\caption{\label{fig:gen} (a) A conceptual diagram of the area $\Omega$ subtended by the closed loop on the HyOPS and definition of the angles $\xi$ and $\zeta$. (b) A ladder chart.}
\end{figure}

\textit{Gauge dependence on $l$}---In general, the Berry phase is gauge invariant. However, we allow the jump from one HyOPS to another HyOPS at the north and the south poles on the HyOPS, so the BP phase may be gauge variant when the gauge depends on $l$. The general solution of Eq.~\eqref{eq:peqH00} is 
\begin{equation}
	\ket{\psi} = e^{\rmi (l'\phi-\delta(z)/2)}R_{l(\phi+\bar{\alpha})} \begin{pmatrix}E_x^0\\ E_y^0e^{\rmi \delta(z)}\end{pmatrix},\label{eq:000_08}
\end{equation}
where $e^{\rmi l\phi}R_{l(\phi+\bar{\alpha})}(E_x^0,E_y^0)^\mathrm{T}$ is the initial state of $\ket{\psi}$ at $z=0$.
An overall phase $\Phi_\text{overall}$ of Eq.~\eqref{eq:000_08} is calculated to be \cite{supp01}
\begin{equation}
	\Phi_\text{overall} = l'\phi-\arg(E_x^0e^{-\rmi\delta/2}\cos l(\phi+\bar{\alpha}) \\- E_y^0e^{i\delta/2}\sin l(\phi+\bar{\alpha})).\label{eq:overall00}
\end{equation}
Because the overall phase does not depend on an initial state \cite{supp01}, we set the initial state to be the north pole of the $l$th HyOPS ($E_x^0 = 1, E_y^0 = \rmi$). In that case, $(R_l,\xi_l,\zeta_l)=(1,\delta,2l\bar{\alpha}-\pi/2)$; thus, the overall phase is rewritten by
\begin{equation}
		\Phi_\text{overall}(\tilde{\bf{S}}_l) = l'\phi+\arctan \frac{\tan(\pi/4+\xi_l/2)}{\tan(l\phi+\zeta_l/2)}.\label{eq:overall00}
\end{equation}
Since Eq.~\eqref{eq:overall00} depends on $l$, which results in the gauge-variant PBP, we express the scalar potential as 
\begin{equation}
\Phi(\tilde{\bm{S}}_l) = \Phi_\text{overall}(\tilde{\bf{S}}_l) + \Phi_\text{indep}(\tilde{\bf{S}}_l),\\
\end{equation}
where $\Phi_\text{indep}$ is an arbitrary function independent of $l$. 
Thus, Eq.~\eqref{eq:vecpot00} can be divided into terms independent of $l$,  ($-\bm{e}_{\zeta_l} (2R_l)^{-1}\tan(\xi_l/2) + \nabla_{\tilde{\bm{S}}_l}\Phi_\text{indep}(\tilde{\bf{S}}_l)$) and the term dependent on $l$, ($\nabla_{\tilde{\bm{S}}_l}\Phi_\text{overall}(\tilde{\bf{S}}_l)$). If the trajectory satisfies the closed loop in the general meaning, the PBP is described by
\begin{eqnarray}
\gamma_\mathrm{Berry} &=& -\frac{\Omega}{2} + \sum_{i=0}^{i_\text{max}}\left [\Phi_\text{overall}(\tilde{\bf{S}}_l)\right ]_{\tilde{\bm{S}}_{l_i}^\mathrm{S}}^{\tilde{\bm{S}}_{l_i}^\mathrm{F}} \label{eq:PBP_01}
\nonumber\\
	&=& -\frac{\Omega}{2} - \frac{\phi}{2}\sum_{i=0}^{i_\text{max}}\left [\left (\cos \xi_{l_i}^\mathrm{F}-\cos\xi_{l_i}^\mathrm{S}\right )l_i\right ],\label{eq:PBP_02}
\end{eqnarray}
where $\Omega$ is the area subtended by the closed loop (Fig.~\ref{fig:gen}(a)). Hence, when the closed loop is on one HyOPS, the PBP is not gauge dependent, but when it travels between HyOPSs, the PBP is gauge dependent. The former and the latter terms of Eq.~\eqref{eq:PBP_02} are  homogeneous and inhomogeneous parts of the PBP, respectively. From the requirement of the closed loop, the inhomogeneous part is quantized by $\phi$, which is illustrated by a ``ladder chart'' (Fig.~\ref{fig:gen}(b)).
This is one of the key results of this letter. 

\begin{figure}[tb]
\includegraphics[width=0.5\textwidth]{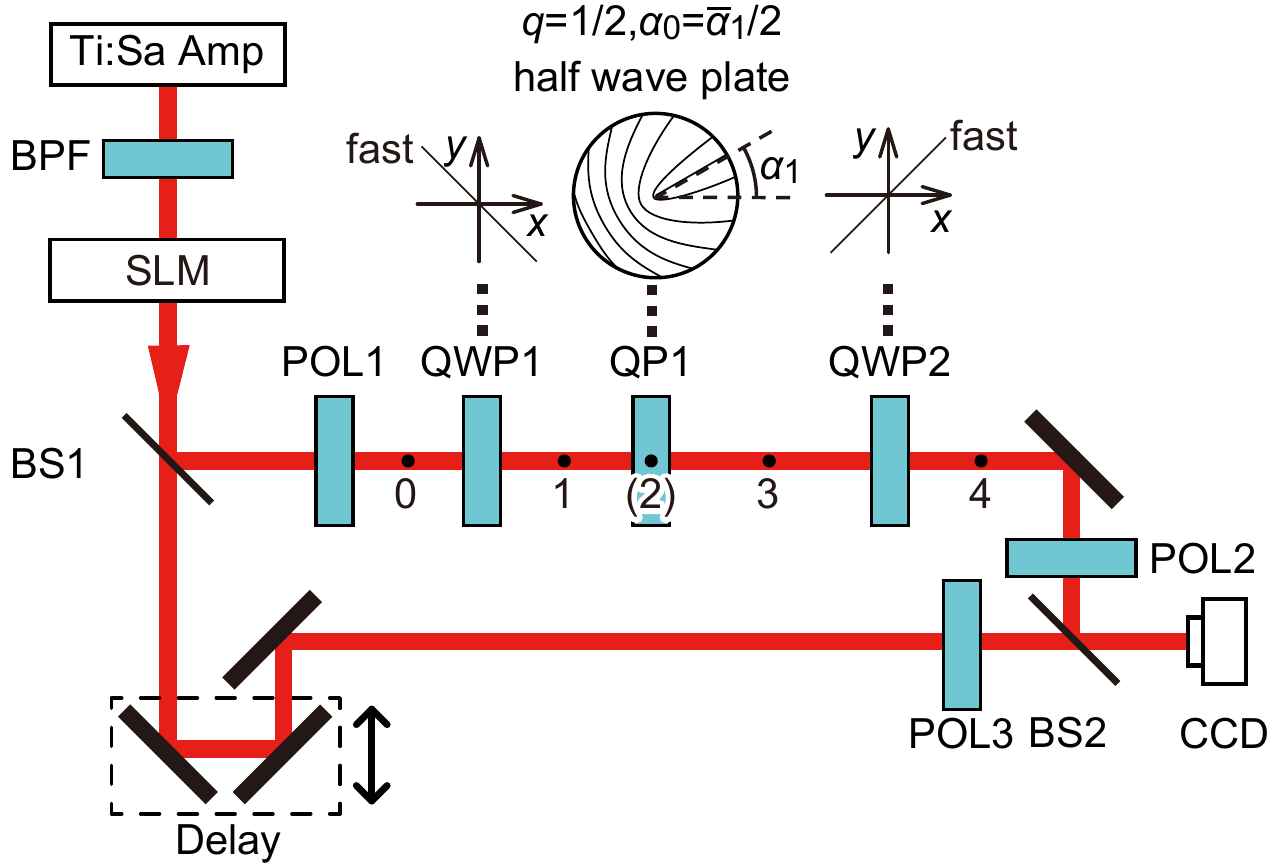}
\caption{\label{fig:setup} The experimental setup for measuring PBP for VVSs, where BPF is a band pass filter (center wavelength, 800\,nm; bandwidth, 10\,nm); SLM is a liquid crystal on silicon spatial light modulator; BS1,2 are $50:50$ non-polarizing beam splitters for ultrafast optics (Thorlabs UFBS5050); POL1,2,3 are polarizers; QWP1,2 are  quarter-wave plates; QP1 is a $q=1/2, \alpha_0=\bar{\alpha}_1/2$ half-wave plate (Photonic Lattice SWP-808); 
Delay is a delay stage; and CCD is a charge coupled device camera. The points are numbered in order to distinguish the intermediate states. Point 2 is located inside QP1.}
\end{figure}

\textit{Experiment.}---Here, we experimentally measured the PBP for VVSs through interferometry analogous to the measurement of the Aharonov--Bohm effect \cite{doi:10.1080/09500348714551321,PhysRev.115.485}. Figure \ref{fig:setup} shows the  experimental setup. 
The light source that we used in this experiment was a Ti:Sapphire laser amplifier (center wavelength, 800\,nm; bandwidth, $\sim$40\,nm; pulse duration, $\sim$25\,fs). The attenuated pulse from the laser amplifier passed through a band pass filter (BPF; center wavelength, 800\,nm; bandwidth, 10\,nm), lengthening its pulse duration to $\sim$120\,fs ($\sim$40\,cycles). We conducted this experiment by use of many-cycle femtosecond pulses \cite{note01}. After the BPF, a spatial light modulator (SLM) system shaped the spatial intensity profile of the pulse into a Gaussian profile. The $x$-polarized (or horizontally polarized) pulse was branched into two beams at a beam splitter (BS1). In the upper branch, the polarization state of light travels on the HyOPSs; in the lower branch, light is directed into the delay line as a reference pulse beam (Fig.~\ref{fig:setup}). Here, BS1 and a second beam splitter (BS2) form a Mach--Zehnder interferometer. In the upper branch, a polarizer (POL1) purified the $x$-polarized state. A quarter-wave plate (QWP1) with the fast axis at $3\pi/4$\,rad to the $x$-axis converted the polarization state into left circularly polarized. After that, the pulse passed through a $q=1/2, \alpha_0=\bar{\alpha}_1/2$ half-wave plate (QP1), and the pulse went through a quarter-wave plate with the fast axis at $\pi/4$\,rad 
to $x$-axis. 
Although the spatial profile is converted into a ``point vortex'' \cite{doi:10.1117/12.910108} or a hypergeometric-Gaussian mode \cite{Karimi:07}  
by a polarization converter QP1, the spatial intensity profile after QP1 was returned to a Gaussian profile on the charge coupled camera (CCD) due to the pair of relay lenses. After the polarizers with polarization axes along the $x$-axis (POL2 and POL3) purified the $x$-polarized states, a beam splitter (BS2) combined the upper and lower beams collinearly and coherently. The delay time was set so that the contrast of the interference was sufficiently high. The spatially interference pattern was captured by  CCD.

\begin{figure}[b]
\includegraphics[width=0.7\textwidth]{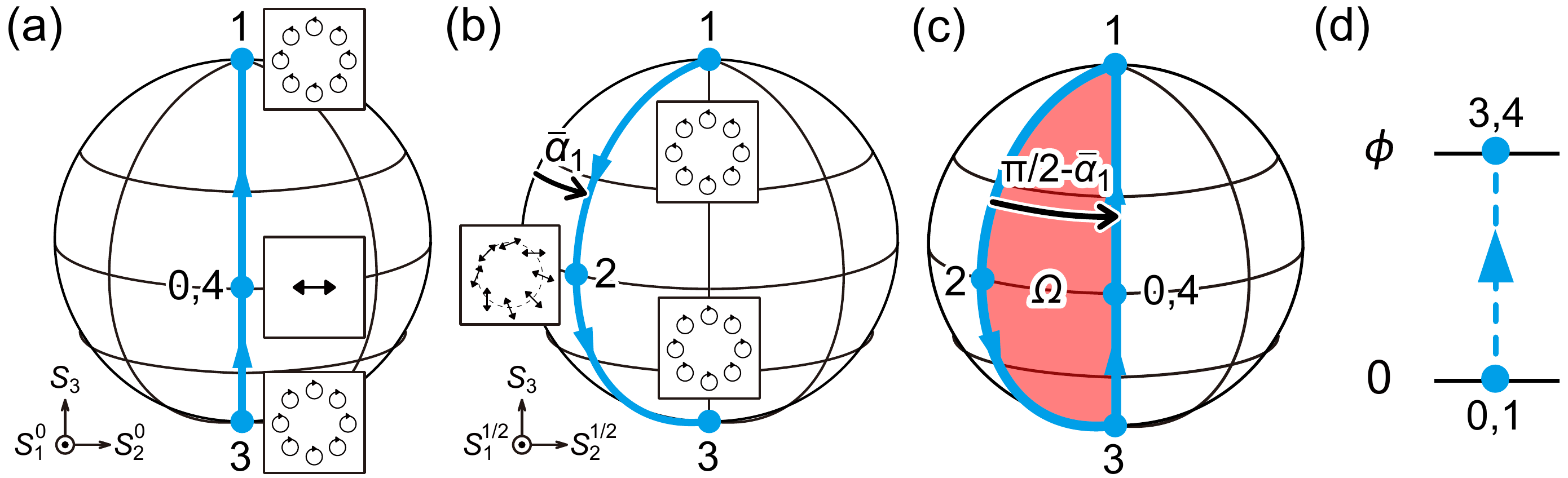}
\caption{\label{fig:di} Paths on the HyOPSs for (a) $l=0$ and (b) $l=1/2$. (c) Superposition of all trajectories on all HyOPSs ((a) and (b)). Here, $\Omega$ is the solid angle of the circuit drawn by the trajectories. State 2 is expedientially illustrated to be  $\tilde{\bm{S}}_{1/2} = \left [\sin(\bar{\alpha}_1-\pi/4),-\cos(\bar{\alpha}_1-\pi/4),0\right ]^\mathrm{T}$. (d) A ladder chart.}
\end{figure}

Figures~\ref{fig:di}(a) and (b) depict the paths of the VVS on the HyOPSs. 
By merging all trajectories into one sphere as shown in Fig.~\ref{fig:di}(c), these paths can form a closed contour, which satisfies the requirement of a ``closed'' loop. From  Eq.~\eqref{eq:PBP_02}, the PBP is calculated to be $-\Omega/2+\phi = \bar{\alpha}_1-\pi/2 + \phi$. Early studies \cite{Biener:02,*Bomzon:02} have mentioned the observation of the inhomogeneous PBP term $\phi$ through a $q=1/2$ half-wave plate, which is known as the space-variant Pancharatnam--Berry phase. However, the uniform PBP term $\bar{\alpha}_1-\pi/2$ has not been explicitly observed. Figure.~\ref{fig:di}(d) is a ladder chart, describing the change in the inhomogeneous part of the PBP.

\begin{figure}[t]
\includegraphics[width=0.5\textwidth]{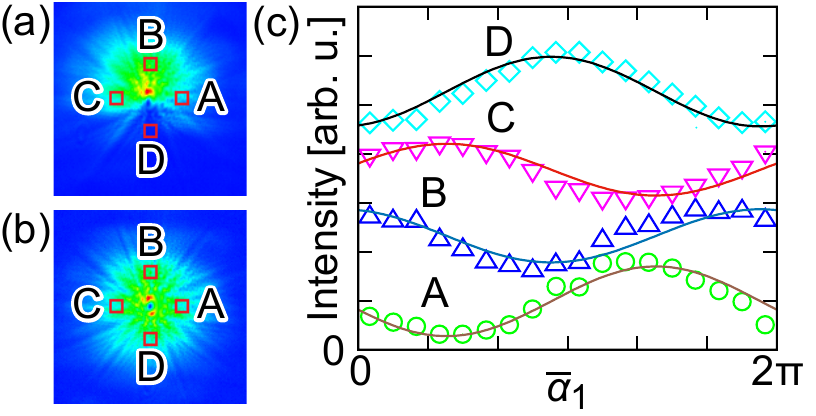}
\caption{\label{fig:results} Experimental results. (a) and (b) are intensity patterns acquired with a CCD camera with and without the reference pulse beam, respectively. Red rectangles in (a) and (b) represent the average areas. (c) Intensity variation by changing the rotation angle $\bar{\alpha}_1$. Curves in (c) are the fitted lines proportional to $\cos(\gamma_\mathrm{Berry}+\delta_\mathrm{pd})+\text{const}$.}
\end{figure}

\textit{Results.}---Figure \ref{fig:results} shows the experimental results. To explore the PBP for VVSs, we measured the intensity of the interference by rotating QP1. The measured intensity patterns are shown in Fig.~\ref{fig:results}(a). The beam center was estimated using the singular point on the intensity pattern without the reference pulses (Fig.~\ref{fig:results}(b)). Areas A, B, C, and D correspond to $\phi = 0, \pi/2, \pi$ and $3\pi/2$, respectively. The intensity is proportional to the function of $\cos(\gamma_\mathrm{Berry}+\delta_\text{pd})+\text{const.}$, where $\delta_\text{pd}$ is the phase difference owing to delay and was experimentally evaluated to be $3.38$\,rad. 
The obtained data in Fig.~\ref{fig:results}(c) are consistent with the above function; thus, we have successfully observed the PBP for VVSs including not only inhomogeneous part but also the homogeneous part. 

\textit{Discussion}---The PBP for VVSs is composed of the homogeneous and inhomogeneous PBP (Eq.~\eqref{eq:PBP_01}). While the former is the same as the PBP for homogeneously polarized states, the latter is unique for VVSs. Thus, we write the general formula of the PBP for VVSs as 
\begin{equation}
	\gamma_\text{Berry}(\phi) = \gamma_\text{Berry,H} + \gamma_\text{Berry,I}(\phi),\label{eq:PBf}
\end{equation}
where $\gamma_\text{PB,H} = -\Omega/2$ and $\gamma_\text{PB,I}(\phi)$, respectively, stand for homogeneous and inhomogeneous PBPs. The former is illustrated by the area subtended by the closed contour on superposed HyOPS (Fig.~\ref{fig:gen}(a)). The latter can be described by the ladder chart in Fig.~\ref{fig:gen}(b), where the intermediate states between the initial and the final states are complicated, when the trajectory satisfies our requirement of the ``closed'' loop.

\begin{figure}[t]
\includegraphics[width=0.5\textwidth]{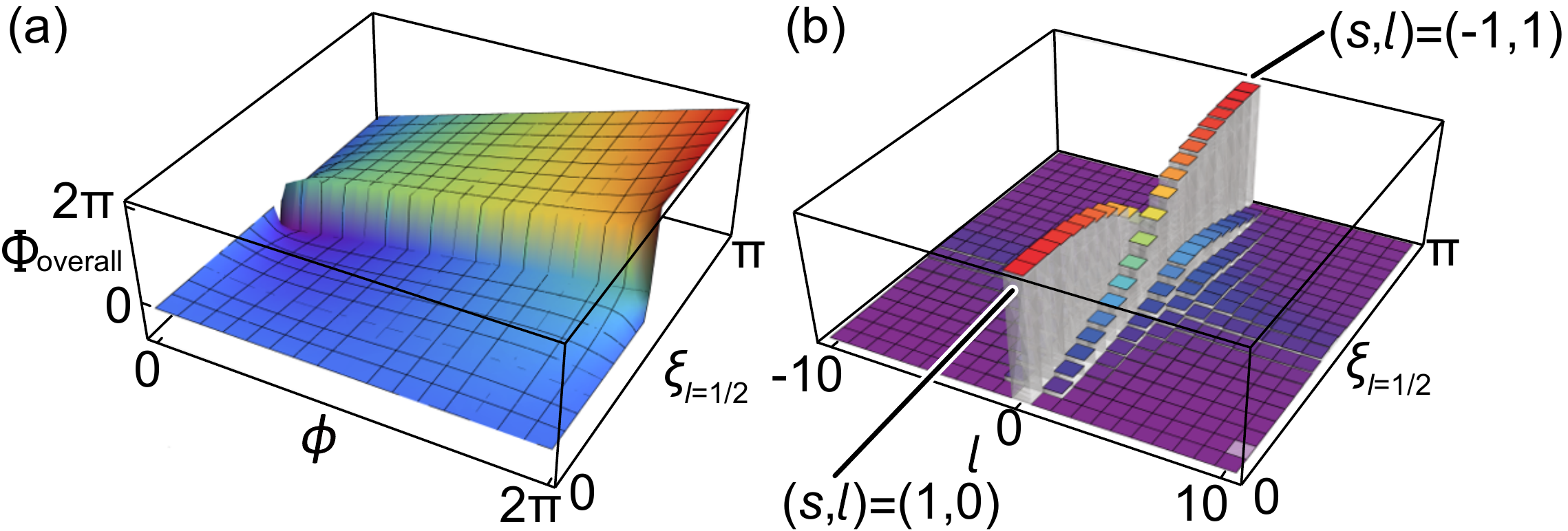}
\caption{\label{fig:pd} (a) Distribution of $\Phi_\text{overall}(\phi,\xi_l)$ and (b) its topological charge power spectrum when $l=1/2$, $\theta_0=\pi/2$, and $l'=1/2$.}
\end{figure}

From Eq.~\eqref{eq:overall00}, in a $q=l$ wave plate, the phase ramp of the inhomogeneous PBP around the beam axis is homogeneous only at the north pole ($\xi_l=0$), the south pole ($\xi_l=\pi$), and the equator ($\xi_l=\pi/2$) of the $l$th HyOPS, but the  distribution along the $\phi$ axis of the inhomogeneous PBP is generally complicated. Figure~\ref{fig:pd} shows the inhomogeneous PBP $\Phi_\text{overall}(\phi,\xi_l)$ and its topological charge distribution 
 (the Fourier power spectrum of $\exp[\rmi\Phi_\text{overall}(\phi,\xi_l)]$ on the $\phi$ axis) for $q=1/2, \alpha=\pi/2$ half-wave plates. 
The transition of VVSs in $q$-plates is interpreted as the optical spin-to-orbital angular momentum conversion 
from $(s,l)\!=\!(\pm 1,l'\mp l)$ to $(\mp 1,l'\pm l)$ ($s$ is the spin angular momentum in units of $\hbar$), which has been conventionally regarded as energy conversion between the left-circularly polarized ($s\!=\!1, \xi_l\!=\!0$) and right-circularly polarized ($s\!=\!-1, \xi_l\!=\!\pi$) states  \cite{PhysRevLett.96.163905,Sakamoto:13}. We further introduce another interpretation of this phenomena: the transition of VVSs in $q$-plates is the adiabatic change of the topological charge spectrum of the inhomogeneous PBP, as shown in Fig.~\ref{fig:pd}(b)

\textit{Conclusion.}---We have introduced the Berry connections of VVSs from the Maxwell--Schr\"odinger equation, and we experimentally verified the PBP obtained from the Berry connections. In contrast to earlier studies, our PBP of VVSs describes an adiabatic change of a VVS on a HiOPS and a HyOPS. We have found that the PBP can be divided into two phases. One phase is the homogeneous PBP $\gamma_\text{PB,H}$, which is essentially the same mathematics as the conventional PBP and is explicitly observed for the first time. The other phase is the inhomogeneous PBP  $\gamma_\text{PB,I}(\phi)$, which is ascribed to the gauge dependence of $l$. We have theoretically detected the adiabatic change of the inhomogeneous PB phase and its topological charge spectrum in polarization converters, which provide another aspect of the optical spin--orbital angular momentum conversion. 
This research was only conducted with $q$-retarders, which confine motions on the HyOPS.   Observations of the PBPs using various optical effects such as Faraday rotation are desired.

Adiabatic manipulation of the quantum coherence of spinor Bose--Einstein condensates (BECs) through the stimulated Raman adiabatic passage (STIRAP) process using optical vortex pulses as pump pulses can be interpreted as a Raman $q$-wave plate for spinor BECs \cite{PhysRevA.77.041601,*PhysRevLett.102.030405,*Schultz:14}. Therefore, the PBP for VVSs has implications for spinor BECs and related other quantum systems. In particular, the homogeneous PBP can be applied to quantum phase gates and precise phase manipulation of macroscopic quantum states.

This work was partially supported by 
a Grant-in-Aid for 
{Scientific Research (B)  (No.~26286056, 2014-2016)}
from 
the Japan Society for the Promotion of Science (JSPS) and CREST, JST.
M.S. acknowledges support from JSPS Research Fellowships (No. 15J00038). 


\appendix

\section{Derivations for the Maxwell--Schr\"{o}dinger equation in the circularly polarized optical vortex basis}

We suppose that the relative permittivity tensor $\hat{\e}$ and the electric field vector $\bm{E}(\bm{r},t)$ are, respectively, described by
\begin{align}
	\hat{\epsilon} =\begin{pmatrix}\e_{xx}&\e_{xy}&0\\\e_{yx}&\e_{yy}&0\\0&0&\e_{zz}\end{pmatrix},\label{eq:S1}\\
	\bm{E}(\bm{r},t) = \tilde{\bm{E}}(\bm{r})e^{-\rmi\omega t}.
\end{align}
From the Maxwell equations, we derive the following two equations \cite{PhysRevLett.80.1888,PhysRevE.81.036602}:
\begin{align}
	\nabla^2\tilde{\bm{E}}+\hat{\e}k^2\tilde{\bm{E}}&=\nabla(\nabla\cdot\tilde{\bm{E}}),\label{eq:S3}\\
	\nabla\cdot(\hat{\e}\tilde{\bm{E}})&=0,\label{eq:S4}
\end{align}
where the dispersion relation in vacuum, $k = \omega/\mathrm{c} \equiv \omega\sqrt{\epsilon_0\mu_0}$, is applied. Here, $\mathrm{c}$, $\epsilon_0$ and $\mu_0$ are the velocity of light in vacuum, the permittivity of vacuum and the permeability of vacuum, respectively.

We express the transverse electric field $\tilde{\bm{E}}$ as
\begin{equation}
	\tilde{\bm{E}}_\bot(\bm{r}) = \begin{pmatrix}\tilde E_x(\bm{r})\\\tilde E_y(\bm{r})\end{pmatrix}
	= e^{\rmi k\sqrt\e z}f(r,z)T^\dag\ket{\phi}
	\equiv e^{\rmi k\sqrt\e z}f(r,z)\frac{1}{\sqrt{2}}\begin{pmatrix}e^{-\rmi l\phi} & e^{\rmi l \phi}\\ \rmi e^{-\rmi l \phi}&-\rmi e^{\rmi l \phi}\end{pmatrix}\begin{pmatrix}\psi_{+,l}(z)\\\psi_{-,l}(z)\end{pmatrix},
\end{equation}
where $\e = (\e_\mathrm{o}+\e_\mathrm{e})$.
Here, we require that $f(r,z)T^\dag$ satisfies the paraxial wave equation:
\begin{equation}
	(\nabla^2_\bot + 2\rmi k\sqrt\e\partial_z)f(r,z)T^\dag = 0.
\end{equation}
From Eq.~\eqref{eq:S1} and Eq.~\eqref{eq:S4}, we derive 
\begin{equation}
	\partial_z \tilde E_z= -\frac{1}{\e_{zz}}\nabla_\bot\cdot\left (\hat{\e}_\bot \tilde{\bm{E}}_\bot\right ),
\end{equation}
thus Eq.~\eqref{eq:S3} is transformed into
\begin{multline}
	\left [(\nabla^2_\bot + 2\rmi k\sqrt{\e}\partial_z)f(r,z)T^\dag \right ]\ket{\phi}+f(r,z)T^\dag\left [(\nabla^2_\bot + 2\rmi k\sqrt{\e}\partial_z) - k^2(\e TT^\dag - T\hat{\e}_\bot T^\dag)\right ]\ket{\phi} \\
	= \nabla_\bot\left [\nabla_\bot\cdot \left ( \hat{1} - \frac{\hat{\e}_\bot}{\e_{zz}} \right )RT^\dag\ket{\phi}) \right ].\label{eq:S8}
\end{multline}

Since $\nabla^2_\bot \ket{\phi} = \bm{0}$, we derive a simplified form of Eq.~\eqref{eq:S8}:
\begin{equation}
	\left [2\rmi k\sqrt{\e}\partial_z - k^2(\e\hat{1} - T\hat{\e}_\bot T^\dag)\right ] \ket{\phi} = \frac{T}{f}\nabla_\bot \left [ \nabla_\bot\cdot \left ( \hat{1} -\frac{\hat{\e}_\bot}{\e_{zz}} \right ) fT^\dag\ket{\phi} \right ].\label{eq:S9}
\end{equation}

Here, we discuss the right-hand side of Eq.~\eqref{eq:S9}. 
If the medium is a $q=l, \alpha_0=l\bar{\alpha}$ retarder, the transverse dielectric tensor is described by
\begin{equation}
	\hat{\e}_\bot = R_{l(\phi+\bar{\alpha})}\begin{pmatrix}\e_\mathrm{o}&0\\0&\e_\mathrm{e}\end{pmatrix}R_{-l(\phi+\bar{\alpha})},
\end{equation}
where $R_\theta$ is a rotation matrix
\begin{equation}
	R_\theta = \begin{pmatrix}\cos\theta & -\sin\theta\\ \sin\theta & \cos\theta\end{pmatrix}.
\end{equation}
The right side of Eq.~\eqref{eq:S9} is regarded as energy conversion term between $\psi_{\pm,l}$, namely the optical spin-orbit coupling \cite{Brasselet:09}. 
 These phenomena are negligible when the collimated beam (or nearly collimated beam) is propagating in a birefringent media \cite{Suzuki:14}. We, therefore, neglect these terms.

Finally, the Maxwell--Schr\"{o}dinger equation in the circularly polarized optical vortex basis is acquired: 
\begin{equation}
	\frac{2\rmi\sqrt{\e}}{k} \partial_z \ket{\phi} = (\e\hat{1} - T\hat{\e}_\bot T^\dag) \ket{\phi}.
\end{equation}

\section{Generic solution for the Maxwell--Schr\"{o}dinger equation of $q$-wave plates and $q$-retarders and its overall phase}

The Maxwell--Schr\"{o}dinger equation for a $q=l$ wave plate is simply given by 
\begin{equation}
	2\rmi\partial_\delta \psi_{\pm,l} = e^{\mp 2\rmi l \bar{\alpha}}\psi_{\mp,l}.
\end{equation}
We obtain the second--order differential equation:
\begin{equation}
	(2i)^2 \partial_\delta^2 \psi_{\pm.l} = \psi_{\pm,l}.
\end{equation}
Thus, the general solution of the Maxwell--Schr\"{o}dinger equation is 
\begin{equation}
	\psi_{\pm,l}(\delta) = A_{\pm+}e^{\rmi\delta/2} + A_{\pm-}e^{-\rmi\delta/2},\label{eq:S16_00}
\end{equation}
where $A_{\pm\pm}$ are constants.
Here, we set the initial condition as follows:
\begin{equation}
	\psi_{\pm,l} (\delta=0) \equiv e^{\mp\rmi\theta_0}A_\pm^0. \label{eq:S17_00}
\end{equation}

Eq.~\eqref{eq:S17_00} describes the relationship between $A_{\pm\pm}$ and $A_\pm^0$:
\begin{align}
	\psi_{\pm,l}(\delta=0) &= A_{\pm+} + A_{\pm-} = +e^{\mp\rmi l\bar{\alpha}}A_\pm^0,\label{eq:S18_00}\\
	\partial_\delta\psi_{\pm,l}|_{\delta=0} &= A_{\pm+} - A_{\pm-} = -e^{\mp\rmi l\bar{\alpha}}A_\mp^0.\label{eq:S19_00}
\end{align}
Using Eq.~\eqref{eq:S18_00} and Eq.~\eqref{eq:S19_00}, the coefficients of Eq.~\eqref{eq:S16_00} are written as
\begin{align}
	A_{\pm+} &= e^{\mp \rmi l\bar{\alpha}}\frac{A_\pm^0-A_\mp^0}{2},\\
	A_{\pm-} &= e^{\mp \rmi l\bar{\alpha}}\frac{A_\pm^0+A_\mp^0}{2}.
\end{align}

Consequently, the general solution is expressed by
\begin{equation}
	\psi_{\pm,l}(\delta) = e^{\mp\rmi l \bar{\alpha}} \left \{ A^0_\pm \cos \left ( \frac{\delta}{2} \right )- \rmi A^0_\mp \sin \left ( \frac{\delta}{2} \right ) \right \},
\end{equation}
and
\begin{align}
	\ket{\psi} &= T^\dag\ket{\phi}\nonumber\\
	&= \frac{e^{il'\phi}e^{-\rmi \delta/2}}{\sqrt{2}}R_{l(\phi+\bar{\alpha})}\begin{pmatrix}A_+^0 + A_-^0 \\ \rmi(A_+^0 - A_-^0)e^{\rmi \delta}\end{pmatrix}\nonumber\\
	&= e^{il'\phi}e^{-\rmi \delta/2}R_{l(\phi+\bar{\alpha})}\begin{pmatrix}E_x^0 \\  E_y^0e^{\rmi \delta}\end{pmatrix},\label{eq:S23_00}
\end{align}
where $e^{\rmi l'\phi}R_{l(\phi+\bar{\alpha})}\begin{pmatrix}E_x^0& E_y^0\end{pmatrix}^\mathrm{T}$ is the initial state vector.

We can transform the vector $\ket{\psi}$ into the vector $\ket{\psi'}$ whose $x$-component is real:
\begin{align}
	\ket{\psi'} &= \exp [-\rmi \Phi_\text{overall} ]\ket{\psi} \\
	&= \begin{pmatrix}|E_x^0e^{-\rmi\delta/2}\cos\theta - E_y^0e^{i\delta/2}\sin\theta|\\(E_x^0e^{-\rmi\delta/2}\sin\theta + E_y^0e^{i\delta/2}\cos\theta)\exp \{-\rmi [l'\phi-\arg(E_x^0e^{-\rmi\delta/2}\cos\theta - E_-^0e^{i\delta/2}\sin\theta)]\}\end{pmatrix},
\end{align}
where
\begin{equation}
	\Phi_\text{overall} = l'\phi-\arg(E_x^0e^{-\rmi\delta/2}\cos\theta - E_-^0e^{i\delta/2}\sin\theta)
\end{equation}
is an overall phase of $\ket{\psi}$ \cite{Visser:10}, and $\theta=l(\phi+\bar{\alpha})$.

Here, we show that the overall phase does not depend on the initial state. Since the overall phase is written in the form of
\begin{equation}
	\Phi_\text{overall} = \int \partial_l (\Phi_\text{overall}) \mathrm{d}l + \text{const.},
\end{equation}
it is sufficient to show that $\partial_l (\Phi_\text{overall})$ is independent of the initial state of
\begin{equation}
	e^{\rmi l'\phi}R_{l(\phi+\theta_0)}\begin{pmatrix}\cos\frac{\alpha}{2}e^{-\rmi\kappa/2}\\\sin\frac{\alpha}{2}e^{\rmi\kappa/2}\end{pmatrix}.
\end{equation}
Here, $\partial_l(\Phi_\text{overall})$ is calculated by
\begin{align}
	\partial_l(\Phi_\text{overall}) &= \partial_l \arg(E_x^0e^{-\rmi\delta/2}\cos\theta - E_-^0e^{i\delta/2}\sin\theta)\\
	&=-\partial_l \arctan \frac{\cos\left( \frac{\alpha}{2} - l\phi - l\bar{\alpha} \right )\sin\left ( \frac{\delta+\kappa}{2} \right )}{\cos\left( \frac{\alpha}{2} + l\phi + l\bar{\alpha} \right )\cos\left ( \frac{\delta+\kappa}{2} \right )}\\
	&= \frac{\phi}{2}\frac{\sin\alpha \sin(\delta+\kappa)}{\cos^2\frac{\alpha}{2}\cos^2(l\phi+l\bar{\alpha})+\sin^2\frac{\alpha}{2}\sin^2(l\phi+l\bar{\alpha})-\frac{1}{2}\sin\alpha\cos(\delta+\kappa)\sin(2l\phi+2l\bar{\alpha})}.
\end{align}
Since the $l$th hybrid-order Stokes parameters are
\begin{align}
	S^l_1 &= \cos (2\bar{\alpha})\cos\alpha - \sin(2\bar{\alpha})\sin\alpha\cos(\delta+\kappa),\\
	S^l_2 &= \sin (2\bar{\alpha})\cos\alpha + \cos(2\bar{\alpha})\sin\alpha \cos(\delta+\kappa),\\
	S^l_3 &= \sin\alpha\sin (\delta+\kappa),
\end{align}
we obtain
\begin{equation}
	\partial_l(\Phi_\text{overall}) = \frac{S_3^l\phi}{1+S_1^l \cos(2l\phi) - S_2^l\sin(2l\phi)}
\end{equation}



\begin{thebibliography}{34}%
\makeatletter
\providecommand \@ifxundefined [1]{%
 \@ifx{#1\undefined}
}%
\providecommand \@ifnum [1]{%
 \ifnum #1\expandafter \@firstoftwo
 \else \expandafter \@secondoftwo
 \fi
}%
\providecommand \@ifx [1]{%
 \ifx #1\expandafter \@firstoftwo
 \else \expandafter \@secondoftwo
 \fi
}%
\providecommand \natexlab [1]{#1}%
\providecommand \enquote  [1]{``#1''}%
\providecommand \bibnamefont  [1]{#1}%
\providecommand \bibfnamefont [1]{#1}%
\providecommand \citenamefont [1]{#1}%
\providecommand \href@noop [0]{\@secondoftwo}%
\providecommand \href [0]{\begingroup \@sanitize@url \@href}%
\providecommand \@href[1]{\@@startlink{#1}\@@href}%
\providecommand \@@href[1]{\endgroup#1\@@endlink}%
\providecommand \@sanitize@url [0]{\catcode `\\12\catcode `\$12\catcode
  `\&12\catcode `\#12\catcode `\^12\catcode `\_12\catcode `\%12\relax}%
\providecommand \@@startlink[1]{}%
\providecommand \@@endlink[0]{}%
\providecommand \url  [0]{\begingroup\@sanitize@url \@url }%
\providecommand \@url [1]{\endgroup\@href {#1}{\urlprefix }}%
\providecommand \urlprefix  [0]{URL }%
\providecommand \Eprint [0]{\href }%
\providecommand \doibase [0]{http://dx.doi.org/}%
\providecommand \selectlanguage [0]{\@gobble}%
\providecommand \bibinfo  [0]{\@secondoftwo}%
\providecommand \bibfield  [0]{\@secondoftwo}%
\providecommand \translation [1]{[#1]}%
\providecommand \BibitemOpen [0]{}%
\providecommand \bibitemStop [0]{}%
\providecommand \bibitemNoStop [0]{.\EOS\space}%
\providecommand \EOS [0]{\spacefactor3000\relax}%
\providecommand \BibitemShut  [1]{\csname bibitem#1\endcsname}%
\let\auto@bib@innerbib\@empty
\bibitem [{\citenamefont {Anandan}(1992)}]{anandan1992geometric}%
  \BibitemOpen
  \bibfield  {author} {\bibinfo {author} {\bibfnamefont {J.}~\bibnamefont
  {Anandan}},\ }\href {http://dx.doi.org/10.1038/360307a0} {\bibfield
  {journal} {\bibinfo  {journal} {Nature}\ }\textbf {\bibinfo {volume} {360}},\
  \bibinfo {pages} {307} (\bibinfo {year} {1992})}\BibitemShut {NoStop}%
\bibitem [{\citenamefont {Pancharatnam}(1956)}]{P1st}%
  \BibitemOpen
  \bibfield  {author} {\bibinfo {author} {\bibfnamefont {S.}~\bibnamefont
  {Pancharatnam}},\ }\href {\doibase 10.1007/BF03046050} {\bibfield  {journal}
  {\bibinfo  {journal} {Proc. Indian Acad. Sci. Sect. A}\ }\textbf {\bibinfo
  {volume} {44}},\ \bibinfo {pages} {247} (\bibinfo {year} {1956})}\BibitemShut
  {NoStop}%
\bibitem [{\citenamefont {Bhandari}\ and\ \citenamefont
  {Samuel}(1988)}]{PhysRevLett.60.1211}%
  \BibitemOpen
  \bibfield  {author} {\bibinfo {author} {\bibfnamefont {R.}~\bibnamefont
  {Bhandari}}\ and\ \bibinfo {author} {\bibfnamefont {J.}~\bibnamefont
  {Samuel}},\ }\href {\doibase 10.1103/PhysRevLett.60.1211} {\bibfield
  {journal} {\bibinfo  {journal} {Phys. Rev. Lett.}\ }\textbf {\bibinfo
  {volume} {60}},\ \bibinfo {pages} {1211} (\bibinfo {year}
  {1988})}\BibitemShut {NoStop}%
\bibitem [{\citenamefont {Chyba}\ \emph {et~al.}(1988)\citenamefont {Chyba},
  \citenamefont {Simon}, \citenamefont {Wang},\ and\ \citenamefont
  {Mandel}}]{Chyba:88}%
  \BibitemOpen
  \bibfield  {author} {\bibinfo {author} {\bibfnamefont {T.~H.}\ \bibnamefont
  {Chyba}}, \bibinfo {author} {\bibfnamefont {R.}~\bibnamefont {Simon}},
  \bibinfo {author} {\bibfnamefont {L.~J.}\ \bibnamefont {Wang}}, \ and\
  \bibinfo {author} {\bibfnamefont {L.}~\bibnamefont {Mandel}},\ }\href
  {\doibase 10.1364/OL.13.000562} {\bibfield  {journal} {\bibinfo  {journal}
  {Opt. Lett.}\ }\textbf {\bibinfo {volume} {13}},\ \bibinfo {pages} {562}
  (\bibinfo {year} {1988})}\BibitemShut {NoStop}%
\bibitem [{\citenamefont {Visser}\ \emph {et~al.}(2010)\citenamefont {Visser},
  \citenamefont {van Dijk}, \citenamefont {Schouten},\ and\ \citenamefont
  {Ubachs}}]{Visser:10}%
  \BibitemOpen
  \bibfield  {author} {\bibinfo {author} {\bibfnamefont {T.~D.}\ \bibnamefont
  {Visser}}, \bibinfo {author} {\bibfnamefont {T.}~\bibnamefont {van Dijk}},
  \bibinfo {author} {\bibfnamefont {H.~F.}\ \bibnamefont {Schouten}}, \ and\
  \bibinfo {author} {\bibfnamefont {W.}~\bibnamefont {Ubachs}},\ }\href
  {\doibase 10.1364/OE.18.010796} {\bibfield  {journal} {\bibinfo  {journal}
  {Opt. Express}\ }\textbf {\bibinfo {volume} {18}},\ \bibinfo {pages} {10796}
  (\bibinfo {year} {2010})}\BibitemShut {NoStop}%
\bibitem [{\citenamefont {Allen}\ \emph {et~al.}(1999)\citenamefont {Allen},
  \citenamefont {Courtial},\ and\ \citenamefont {Padgett}}]{PhysRevE.60.7497}%
  \BibitemOpen
  \bibfield  {author} {\bibinfo {author} {\bibfnamefont {L.}~\bibnamefont
  {Allen}}, \bibinfo {author} {\bibfnamefont {J.}~\bibnamefont {Courtial}}, \
  and\ \bibinfo {author} {\bibfnamefont {M.~J.}\ \bibnamefont {Padgett}},\
  }\href {\doibase 10.1103/PhysRevE.60.7497} {\bibfield  {journal} {\bibinfo
  {journal} {Phys. Rev. E}\ }\textbf {\bibinfo {volume} {60}},\ \bibinfo
  {pages} {7497} (\bibinfo {year} {1999})}\BibitemShut {NoStop}%
\bibitem [{\citenamefont {Milione}\ \emph {et~al.}(2012)\citenamefont
  {Milione}, \citenamefont {Evans}, \citenamefont {Nolan},\ and\ \citenamefont
  {Alfano}}]{PhysRevLett.108.190401}%
  \BibitemOpen
  \bibfield  {author} {\bibinfo {author} {\bibfnamefont {G.}~\bibnamefont
  {Milione}}, \bibinfo {author} {\bibfnamefont {S.}~\bibnamefont {Evans}},
  \bibinfo {author} {\bibfnamefont {D.~A.}\ \bibnamefont {Nolan}}, \ and\
  \bibinfo {author} {\bibfnamefont {R.~R.}\ \bibnamefont {Alfano}},\ }\href
  {\doibase 10.1103/PhysRevLett.108.190401} {\bibfield  {journal} {\bibinfo
  {journal} {Phys. Rev. Lett.}\ }\textbf {\bibinfo {volume} {108}},\ \bibinfo
  {pages} {190401} (\bibinfo {year} {2012})}\BibitemShut {NoStop}%
\bibitem [{\citenamefont {Toyoda}\ \emph {et~al.}(2013)\citenamefont {Toyoda},
  \citenamefont {Takahashi}, \citenamefont {Takizawa}, \citenamefont
  {Tokizane}, \citenamefont {Miyamoto}, \citenamefont {Morita},\ and\
  \citenamefont {Omatsu}}]{PhysRevLett.110.143603}%
  \BibitemOpen
  \bibfield  {author} {\bibinfo {author} {\bibfnamefont {K.}~\bibnamefont
  {Toyoda}}, \bibinfo {author} {\bibfnamefont {F.}~\bibnamefont {Takahashi}},
  \bibinfo {author} {\bibfnamefont {S.}~\bibnamefont {Takizawa}}, \bibinfo
  {author} {\bibfnamefont {Y.}~\bibnamefont {Tokizane}}, \bibinfo {author}
  {\bibfnamefont {K.}~\bibnamefont {Miyamoto}}, \bibinfo {author}
  {\bibfnamefont {R.}~\bibnamefont {Morita}}, \ and\ \bibinfo {author}
  {\bibfnamefont {T.}~\bibnamefont {Omatsu}},\ }\href {\doibase
  10.1103/PhysRevLett.110.143603} {\bibfield  {journal} {\bibinfo  {journal}
  {Phys. Rev. Lett.}\ }\textbf {\bibinfo {volume} {110}},\ \bibinfo {pages}
  {143603} (\bibinfo {year} {2013})}\BibitemShut {NoStop}%
\bibitem [{\citenamefont {Parigi}\ \emph {et~al.}(2015)\citenamefont {Parigi},
  \citenamefont {D'Ambrosio}, \citenamefont {Arnold}, \citenamefont {Marrucci},
  \citenamefont {Sciarrino},\ and\ \citenamefont {Laurat}}]{parigi2015storage}%
  \BibitemOpen
  \bibfield  {author} {\bibinfo {author} {\bibfnamefont {V.}~\bibnamefont
  {Parigi}}, \bibinfo {author} {\bibfnamefont {V.}~\bibnamefont {D'Ambrosio}},
  \bibinfo {author} {\bibfnamefont {C.}~\bibnamefont {Arnold}}, \bibinfo
  {author} {\bibfnamefont {L.}~\bibnamefont {Marrucci}}, \bibinfo {author}
  {\bibfnamefont {F.}~\bibnamefont {Sciarrino}}, \ and\ \bibinfo {author}
  {\bibfnamefont {J.}~\bibnamefont {Laurat}},\ }\href {\doibase
  10.1038/ncomms8706} {\bibfield  {journal} {\bibinfo  {journal} {Nat.
  Commun.}\ }\textbf {\bibinfo {volume} {6}},\ \bibinfo {pages} {7706}
  (\bibinfo {year} {2015})}\BibitemShut {NoStop}%
\bibitem [{\citenamefont {Maurer}\ \emph {et~al.}(2007)\citenamefont {Maurer},
  \citenamefont {Jesacher}, \citenamefont {F\"{u}rhapter}, \citenamefont
  {Bernet},\ and\ \citenamefont {Ritsch-Marte}}]{1367-2630-9-3-078}%
  \BibitemOpen
  \bibfield  {author} {\bibinfo {author} {\bibfnamefont {C.}~\bibnamefont
  {Maurer}}, \bibinfo {author} {\bibfnamefont {A.}~\bibnamefont {Jesacher}},
  \bibinfo {author} {\bibfnamefont {S.}~\bibnamefont {F\"{u}rhapter}}, \bibinfo
  {author} {\bibfnamefont {S.}~\bibnamefont {Bernet}}, \ and\ \bibinfo {author}
  {\bibfnamefont {M.}~\bibnamefont {Ritsch-Marte}},\ }\href
  {http://stacks.iop.org/1367-2630/9/i=3/a=078} {\bibfield  {journal} {\bibinfo
   {journal} {New Journal of Physics}\ }\textbf {\bibinfo {volume} {9}},\
  \bibinfo {pages} {78} (\bibinfo {year} {2007})}\BibitemShut {NoStop}%
\bibitem [{\citenamefont {Milione}\ \emph {et~al.}(2011)\citenamefont
  {Milione}, \citenamefont {Sztul}, \citenamefont {Nolan},\ and\ \citenamefont
  {Alfano}}]{PhysRevLett.107.053601}%
  \BibitemOpen
  \bibfield  {author} {\bibinfo {author} {\bibfnamefont {G.}~\bibnamefont
  {Milione}}, \bibinfo {author} {\bibfnamefont {H.~I.}\ \bibnamefont {Sztul}},
  \bibinfo {author} {\bibfnamefont {D.~A.}\ \bibnamefont {Nolan}}, \ and\
  \bibinfo {author} {\bibfnamefont {R.~R.}\ \bibnamefont {Alfano}},\ }\href
  {\doibase 10.1103/PhysRevLett.107.053601} {\bibfield  {journal} {\bibinfo
  {journal} {Phys. Rev. Lett.}\ }\textbf {\bibinfo {volume} {107}},\ \bibinfo
  {pages} {053601} (\bibinfo {year} {2011})}\BibitemShut {NoStop}%
\bibitem [{\citenamefont {Holleczek}\ \emph {et~al.}(2011)\citenamefont
  {Holleczek}, \citenamefont {Aiello}, \citenamefont {Gabriel}, \citenamefont
  {Marquardt},\ and\ \citenamefont {Leuchs}}]{Holleczek:11}%
  \BibitemOpen
  \bibfield  {author} {\bibinfo {author} {\bibfnamefont {A.}~\bibnamefont
  {Holleczek}}, \bibinfo {author} {\bibfnamefont {A.}~\bibnamefont {Aiello}},
  \bibinfo {author} {\bibfnamefont {C.}~\bibnamefont {Gabriel}}, \bibinfo
  {author} {\bibfnamefont {C.}~\bibnamefont {Marquardt}}, \ and\ \bibinfo
  {author} {\bibfnamefont {G.}~\bibnamefont {Leuchs}},\ }\href {\doibase
  10.1364/OE.19.009714} {\bibfield  {journal} {\bibinfo  {journal} {Opt.
  Express}\ }\textbf {\bibinfo {volume} {19}},\ \bibinfo {pages} {9714}
  (\bibinfo {year} {2011})}\BibitemShut {NoStop}%
\bibitem [{\citenamefont {Yi}\ \emph {et~al.}(2015)\citenamefont {Yi},
  \citenamefont {Liu}, \citenamefont {Ling}, \citenamefont {Zhou},
  \citenamefont {Ke}, \citenamefont {Luo}, \citenamefont {Wen},\ and\
  \citenamefont {Fan}}]{PhysRevA.91.023801}%
  \BibitemOpen
  \bibfield  {author} {\bibinfo {author} {\bibfnamefont {X.}~\bibnamefont
  {Yi}}, \bibinfo {author} {\bibfnamefont {Y.}~\bibnamefont {Liu}}, \bibinfo
  {author} {\bibfnamefont {X.}~\bibnamefont {Ling}}, \bibinfo {author}
  {\bibfnamefont {X.}~\bibnamefont {Zhou}}, \bibinfo {author} {\bibfnamefont
  {Y.}~\bibnamefont {Ke}}, \bibinfo {author} {\bibfnamefont {H.}~\bibnamefont
  {Luo}}, \bibinfo {author} {\bibfnamefont {S.}~\bibnamefont {Wen}}, \ and\
  \bibinfo {author} {\bibfnamefont {D.}~\bibnamefont {Fan}},\ }\href {\doibase
  10.1103/PhysRevA.91.023801} {\bibfield  {journal} {\bibinfo  {journal} {Phys.
  Rev. A}\ }\textbf {\bibinfo {volume} {91}},\ \bibinfo {pages} {023801}
  (\bibinfo {year} {2015})}\BibitemShut {NoStop}%
\bibitem [{\citenamefont {Berry}(1984)}]{Berry45}%
  \BibitemOpen
  \bibfield  {author} {\bibinfo {author} {\bibfnamefont {M.~V.}\ \bibnamefont
  {Berry}},\ }\href {\doibase 10.1098/rspa.1984.0023} {\bibfield  {journal}
  {\bibinfo  {journal} {Proc. R. Soc., Lond. A}\ }\textbf {\bibinfo {volume}
  {392}},\ \bibinfo {pages} {45} (\bibinfo {year} {1984})}\BibitemShut
  {NoStop}%
\bibitem [{\citenamefont {Suzuki}\ \emph {et~al.}()\citenamefont {Suzuki},
  \citenamefont {Yamane}, \citenamefont {Oka}, \citenamefont {Toda},\ and\
  \citenamefont {Morita}}]{Suzuk_prlcomment}%
  \BibitemOpen
  \bibfield  {author} {\bibinfo {author} {\bibfnamefont {M.}~\bibnamefont
  {Suzuki}}, \bibinfo {author} {\bibfnamefont {K.}~\bibnamefont {Yamane}},
  \bibinfo {author} {\bibfnamefont {K.}~\bibnamefont {Oka}}, \bibinfo {author}
  {\bibfnamefont {Y.}~\bibnamefont {Toda}}, \ and\ \bibinfo {author}
  {\bibfnamefont {R.}~\bibnamefont {Morita}},\ }\href@noop {} {\enquote
  {\bibinfo {title} {Comment on `{H}igher order {P}ancharatnam-{B}erry phase
  and the angular momentum of light'},}\ }\BibitemShut {NoStop}%
\bibitem [{\citenamefont {Chiao}\ and\ \citenamefont
  {Wu}(1986)}]{PhysRevLett.57.933}%
  \BibitemOpen
  \bibfield  {author} {\bibinfo {author} {\bibfnamefont {R.~Y.}\ \bibnamefont
  {Chiao}}\ and\ \bibinfo {author} {\bibfnamefont {Y.-S.}\ \bibnamefont {Wu}},\
  }\href {\doibase 10.1103/PhysRevLett.57.933} {\bibfield  {journal} {\bibinfo
  {journal} {Phys. Rev. Lett.}\ }\textbf {\bibinfo {volume} {57}},\ \bibinfo
  {pages} {933} (\bibinfo {year} {1986})}\BibitemShut {NoStop}%
\bibitem [{\citenamefont {Berry}(1987)}]{doi:10.1080/09500348714551321}%
  \BibitemOpen
  \bibfield  {author} {\bibinfo {author} {\bibfnamefont {M.~V.}\ \bibnamefont
  {Berry}},\ }\href {\doibase 10.1080/09500348714551321} {\bibfield  {journal}
  {\bibinfo  {journal} {Journal of Modern Optics}\ }\textbf {\bibinfo {volume}
  {34}},\ \bibinfo {pages} {1401} (\bibinfo {year} {1987})}\BibitemShut
  {NoStop}%
\bibitem [{\citenamefont {Marrucci}\ \emph {et~al.}(2006)\citenamefont
  {Marrucci}, \citenamefont {Manzo},\ and\ \citenamefont
  {Paparo}}]{PhysRevLett.96.163905}%
  \BibitemOpen
  \bibfield  {author} {\bibinfo {author} {\bibfnamefont {L.}~\bibnamefont
  {Marrucci}}, \bibinfo {author} {\bibfnamefont {C.}~\bibnamefont {Manzo}}, \
  and\ \bibinfo {author} {\bibfnamefont {D.}~\bibnamefont {Paparo}},\ }\href
  {\doibase 10.1103/PhysRevLett.96.163905} {\bibfield  {journal} {\bibinfo
  {journal} {Phys. Rev. Lett.}\ }\textbf {\bibinfo {volume} {96}},\ \bibinfo
  {pages} {163905} (\bibinfo {year} {2006})}\BibitemShut {NoStop}%
\bibitem [{\citenamefont {Kuratsuji}\ and\ \citenamefont
  {Kakigi}(1998)}]{PhysRevLett.80.1888}%
  \BibitemOpen
  \bibfield  {author} {\bibinfo {author} {\bibfnamefont {H.}~\bibnamefont
  {Kuratsuji}}\ and\ \bibinfo {author} {\bibfnamefont {S.}~\bibnamefont
  {Kakigi}},\ }\href {\doibase 10.1103/PhysRevLett.80.1888} {\bibfield
  {journal} {\bibinfo  {journal} {Phys. Rev. Lett.}\ }\textbf {\bibinfo
  {volume} {80}},\ \bibinfo {pages} {1888} (\bibinfo {year}
  {1998})}\BibitemShut {NoStop}%
\bibitem [{\citenamefont {Botet}\ and\ \citenamefont
  {Kuratsuji}(2010)}]{PhysRevE.81.036602}%
  \BibitemOpen
  \bibfield  {author} {\bibinfo {author} {\bibfnamefont {R.}~\bibnamefont
  {Botet}}\ and\ \bibinfo {author} {\bibfnamefont {H.}~\bibnamefont
  {Kuratsuji}},\ }\href {\doibase 10.1103/PhysRevE.81.036602} {\bibfield
  {journal} {\bibinfo  {journal} {Phys. Rev. E}\ }\textbf {\bibinfo {volume}
  {81}},\ \bibinfo {pages} {036602} (\bibinfo {year} {2010})}\BibitemShut
  {NoStop}%
\bibitem [{sup()}]{supp01}%
  \BibitemOpen
  \href@noop {} {}\bibinfo {note} {See supplemental material at [URL] for the
  detailed derivations of the Maxwell-Schor\"{o}dinger equation
  Eq.~\eqref{eq:peqH00} and the detailed discussion of the overall phase
  Eq.~\eqref{eq:overall00}.}\BibitemShut {Stop}%
\bibitem [{\citenamefont {Holstein}(1989)}]{Holstein89}%
  \BibitemOpen
  \bibfield  {author} {\bibinfo {author} {\bibfnamefont {B.~R.}\ \bibnamefont
  {Holstein}},\ }\href {\doibase 10.1119/1.15793} {\bibfield  {journal}
  {\bibinfo  {journal} {American Journal of Physics}\ }\textbf {\bibinfo
  {volume} {57}},\ \bibinfo {pages} {1079} (\bibinfo {year}
  {1989})}\BibitemShut {NoStop}%
\bibitem [{\citenamefont {Aharonov}\ and\ \citenamefont
  {Bohm}(1959)}]{PhysRev.115.485}%
  \BibitemOpen
  \bibfield  {author} {\bibinfo {author} {\bibfnamefont {Y.}~\bibnamefont
  {Aharonov}}\ and\ \bibinfo {author} {\bibfnamefont {D.}~\bibnamefont
  {Bohm}},\ }\href {\doibase 10.1103/PhysRev.115.485} {\bibfield  {journal}
  {\bibinfo  {journal} {Phys. Rev.}\ }\textbf {\bibinfo {volume} {115}},\
  \bibinfo {pages} {485} (\bibinfo {year} {1959})}\BibitemShut {NoStop}%
\bibitem [{not()}]{note01}%
  \BibitemOpen
  \href@noop {} {}\bibinfo {note} {Similar experiments can be performed by CW
  lasers}\BibitemShut {NoStop}%
\bibitem [{\citenamefont {Sakamoto}\ \emph {et~al.}(2012)\citenamefont
  {Sakamoto}, \citenamefont {Oka},\ and\ \citenamefont
  {Morita}}]{doi:10.1117/12.910108}%
  \BibitemOpen
  \bibfield  {author} {\bibinfo {author} {\bibfnamefont {M.}~\bibnamefont
  {Sakamoto}}, \bibinfo {author} {\bibfnamefont {K.}~\bibnamefont {Oka}}, \
  and\ \bibinfo {author} {\bibfnamefont {R.}~\bibnamefont {Morita}},\ }\href
  {\doibase 10.1117/12.910108} {\bibfield  {journal} {\bibinfo  {journal}
  {Proc. SPIE}\ }\textbf {\bibinfo {volume} {8274}},\ \bibinfo {pages} {827414}
  (\bibinfo {year} {2012})}\BibitemShut {NoStop}%
\bibitem [{\citenamefont {Karimi}\ \emph {et~al.}(2007)\citenamefont {Karimi},
  \citenamefont {Zito}, \citenamefont {Piccirillo}, \citenamefont {Marrucci},\
  and\ \citenamefont {Santamato}}]{Karimi:07}%
  \BibitemOpen
  \bibfield  {author} {\bibinfo {author} {\bibfnamefont {E.}~\bibnamefont
  {Karimi}}, \bibinfo {author} {\bibfnamefont {G.}~\bibnamefont {Zito}},
  \bibinfo {author} {\bibfnamefont {B.}~\bibnamefont {Piccirillo}}, \bibinfo
  {author} {\bibfnamefont {L.}~\bibnamefont {Marrucci}}, \ and\ \bibinfo
  {author} {\bibfnamefont {E.}~\bibnamefont {Santamato}},\ }\href {\doibase
  10.1364/OL.32.003053} {\bibfield  {journal} {\bibinfo  {journal} {Opt.
  Lett.}\ }\textbf {\bibinfo {volume} {32}},\ \bibinfo {pages} {3053} (\bibinfo
  {year} {2007})}\BibitemShut {NoStop}%
\bibitem [{\citenamefont {Biener}\ \emph {et~al.}(2002)\citenamefont {Biener},
  \citenamefont {Niv}, \citenamefont {Kleiner},\ and\ \citenamefont
  {Hasman}}]{Biener:02}%
  \BibitemOpen
  \bibfield  {author} {\bibinfo {author} {\bibfnamefont {G.}~\bibnamefont
  {Biener}}, \bibinfo {author} {\bibfnamefont {A.}~\bibnamefont {Niv}},
  \bibinfo {author} {\bibfnamefont {V.}~\bibnamefont {Kleiner}}, \ and\
  \bibinfo {author} {\bibfnamefont {E.}~\bibnamefont {Hasman}},\ }\href
  {\doibase 10.1364/OL.27.001875} {\bibfield  {journal} {\bibinfo  {journal}
  {Opt. Lett.}\ }\textbf {\bibinfo {volume} {27}},\ \bibinfo {pages} {1875}
  (\bibinfo {year} {2002})}\BibitemShut {NoStop}%
\bibitem [{\citenamefont {Bomzon}\ \emph {et~al.}(2002)\citenamefont {Bomzon},
  \citenamefont {Biener}, \citenamefont {Kleiner},\ and\ \citenamefont
  {Hasman}}]{Bomzon:02}%
  \BibitemOpen
  \bibfield  {author} {\bibinfo {author} {\bibfnamefont {Z.}~\bibnamefont
  {Bomzon}}, \bibinfo {author} {\bibfnamefont {G.}~\bibnamefont {Biener}},
  \bibinfo {author} {\bibfnamefont {V.}~\bibnamefont {Kleiner}}, \ and\
  \bibinfo {author} {\bibfnamefont {E.}~\bibnamefont {Hasman}},\ }\href
  {\doibase 10.1364/OL.27.001141} {\bibfield  {journal} {\bibinfo  {journal}
  {Opt. Lett.}\ }\textbf {\bibinfo {volume} {27}},\ \bibinfo {pages} {1141}
  (\bibinfo {year} {2002})}\BibitemShut {NoStop}%
\bibitem [{\citenamefont {Sakamoto}\ \emph {et~al.}(2013)\citenamefont
  {Sakamoto}, \citenamefont {Oka}, \citenamefont {Morita},\ and\ \citenamefont
  {Murakami}}]{Sakamoto:13}%
  \BibitemOpen
  \bibfield  {author} {\bibinfo {author} {\bibfnamefont {M.}~\bibnamefont
  {Sakamoto}}, \bibinfo {author} {\bibfnamefont {K.}~\bibnamefont {Oka}},
  \bibinfo {author} {\bibfnamefont {R.}~\bibnamefont {Morita}}, \ and\ \bibinfo
  {author} {\bibfnamefont {N.}~\bibnamefont {Murakami}},\ }\href {\doibase
  10.1364/OL.38.003661} {\bibfield  {journal} {\bibinfo  {journal} {Opt.
  Lett.}\ }\textbf {\bibinfo {volume} {38}},\ \bibinfo {pages} {3661} (\bibinfo
  {year} {2013})}\BibitemShut {NoStop}%
\bibitem [{\citenamefont {Wright}\ \emph {et~al.}(2008)\citenamefont {Wright},
  \citenamefont {Leslie},\ and\ \citenamefont {Bigelow}}]{PhysRevA.77.041601}%
  \BibitemOpen
  \bibfield  {author} {\bibinfo {author} {\bibfnamefont {K.~C.}\ \bibnamefont
  {Wright}}, \bibinfo {author} {\bibfnamefont {L.~S.}\ \bibnamefont {Leslie}},
  \ and\ \bibinfo {author} {\bibfnamefont {N.~P.}\ \bibnamefont {Bigelow}},\
  }\href {\doibase 10.1103/PhysRevA.77.041601} {\bibfield  {journal} {\bibinfo
  {journal} {Phys. Rev. A}\ }\textbf {\bibinfo {volume} {77}},\ \bibinfo
  {pages} {041601} (\bibinfo {year} {2008})}\BibitemShut {NoStop}%
\bibitem [{\citenamefont {Wright}\ \emph {et~al.}(2009)\citenamefont {Wright},
  \citenamefont {Leslie}, \citenamefont {Hansen},\ and\ \citenamefont
  {Bigelow}}]{PhysRevLett.102.030405}%
  \BibitemOpen
  \bibfield  {author} {\bibinfo {author} {\bibfnamefont {K.~C.}\ \bibnamefont
  {Wright}}, \bibinfo {author} {\bibfnamefont {L.~S.}\ \bibnamefont {Leslie}},
  \bibinfo {author} {\bibfnamefont {A.}~\bibnamefont {Hansen}}, \ and\ \bibinfo
  {author} {\bibfnamefont {N.~P.}\ \bibnamefont {Bigelow}},\ }\href {\doibase
  10.1103/PhysRevLett.102.030405} {\bibfield  {journal} {\bibinfo  {journal}
  {Phys. Rev. Lett.}\ }\textbf {\bibinfo {volume} {102}},\ \bibinfo {pages}
  {030405} (\bibinfo {year} {2009})}\BibitemShut {NoStop}%
\bibitem [{\citenamefont {Schultz}\ \emph {et~al.}(2014)\citenamefont
  {Schultz}, \citenamefont {Hansen},\ and\ \citenamefont
  {Bigelow}}]{Schultz:14}%
  \BibitemOpen
  \bibfield  {author} {\bibinfo {author} {\bibfnamefont {J.~T.}\ \bibnamefont
  {Schultz}}, \bibinfo {author} {\bibfnamefont {A.}~\bibnamefont {Hansen}}, \
  and\ \bibinfo {author} {\bibfnamefont {N.~P.}\ \bibnamefont {Bigelow}},\
  }\href {\doibase 10.1364/OL.39.004271} {\bibfield  {journal} {\bibinfo
  {journal} {Opt. Lett.}\ }\textbf {\bibinfo {volume} {39}},\ \bibinfo {pages}
  {4271} (\bibinfo {year} {2014})}\BibitemShut {NoStop}%
\bibitem [{\citenamefont {Brasselet}\ \emph {et~al.}(2009)\citenamefont
  {Brasselet}, \citenamefont {Izdebskaya}, \citenamefont {Shvedov},
  \citenamefont {Desyatnikov}, \citenamefont {Krolikowski},\ and\ \citenamefont
  {Kivshar}}]{Brasselet:09}%
  \BibitemOpen
  \bibfield  {author} {\bibinfo {author} {\bibfnamefont {E.}~\bibnamefont
  {Brasselet}}, \bibinfo {author} {\bibfnamefont {Y.}~\bibnamefont
  {Izdebskaya}}, \bibinfo {author} {\bibfnamefont {V.}~\bibnamefont {Shvedov}},
  \bibinfo {author} {\bibfnamefont {A.~S.}\ \bibnamefont {Desyatnikov}},
  \bibinfo {author} {\bibfnamefont {W.}~\bibnamefont {Krolikowski}}, \ and\
  \bibinfo {author} {\bibfnamefont {Y.~S.}\ \bibnamefont {Kivshar}},\ }\href
  {\doibase 10.1364/OL.34.001021} {\bibfield  {journal} {\bibinfo  {journal}
  {Opt. Lett.}\ }\textbf {\bibinfo {volume} {34}},\ \bibinfo {pages} {1021}
  (\bibinfo {year} {2009})}\BibitemShut {NoStop}%
\bibitem [{\citenamefont {Suzuki}\ \emph {et~al.}(2014)\citenamefont {Suzuki},
  \citenamefont {Yamane}, \citenamefont {Oka}, \citenamefont {Toda},\ and\
  \citenamefont {Morita}}]{Suzuki:14}%
  \BibitemOpen
  \bibfield  {author} {\bibinfo {author} {\bibfnamefont {M.}~\bibnamefont
  {Suzuki}}, \bibinfo {author} {\bibfnamefont {K.}~\bibnamefont {Yamane}},
  \bibinfo {author} {\bibfnamefont {K.}~\bibnamefont {Oka}}, \bibinfo {author}
  {\bibfnamefont {Y.}~\bibnamefont {Toda}}, \ and\ \bibinfo {author}
  {\bibfnamefont {R.}~\bibnamefont {Morita}},\ }\href {\doibase
  10.1364/OE.22.016903} {\bibfield  {journal} {\bibinfo  {journal} {Opt.
  Express}\ }\textbf {\bibinfo {volume} {22}},\ \bibinfo {pages} {16903}
  (\bibinfo {year} {2014})}\BibitemShut {NoStop}%
\end{thebibliography}
\providecommand{\noopsort}[1]{}\providecommand{\singleletter}[1]{#1}%

\end{document}